\newcommand{\beq}{\begin{equation}}
\newcommand{\eeq}{\end{equation}}
\newcommand{\bea}{\begin{eqnarray}}
\newcommand{\eea}{\end{eqnarray}}

\newcommand{\gsim}{\lower.7ex\hbox{$\;\stackrel{\textstyle>}{\sim}\;$}}
\newcommand{\lsim}{\lower.7ex\hbox{$\;\stackrel{\textstyle<}{\sim}\;$}}

\documentclass[aps,prev,preprintnumbers,floatfix,nofootinbib,11pt]{revtex4-1}
\usepackage{graphicx}
\usepackage{epstopdf}
\usepackage{mathrsfs}
\usepackage{amssymb}
\usepackage{verbatim}
\usepackage{xcolor}
\usepackage{multirow}
\usepackage{amsmath}
\usepackage{float}
\usepackage[normalem]{ulem}
\usepackage[lofdepth,lotdepth,caption=false]{subfig}

\def\stacksymbols #1#2#3#4{\def\theguybelow{#2}
    \def\vp{\lower#3pt}
    \def\sp{\baselineskip0pt\lineskip#4pt}
    \mathrel{\mathpalette\intermediary#1}}

\def\intermediary#1#2{\vp\vbox{\sp
     \everycr={}\tabskip0pt
     \halign{$\mathsurround0pt#1\hfil##\hfil$\crcr#2\crcr
              \theguybelow\crcr}}}

\def\be{\begin{equation}}
\def\ee{\end{equation}}
\def\bea{\begin{eqnarray}}
\def\eea{\end{eqnarray}}
\begin{document}

\vspace*{1mm}

\title{Excess of Tau events at SND@LHC, FASER$\nu$ and  FASER$\nu$2 }
\author{Saeed Ansarifard}
\email{ansarifard@ipm.ir}
\author{Yasaman Farzan}
\email{yasaman@theory.ipm.ac.ir}

\vspace{0.2cm}

\affiliation{
	${}^a$School of physics, Institute for Research in Fundamental Sciences (IPM),\\
	P.O. Box 19395-5531, Tehran, Iran
}

\begin{abstract}  
	 During the run III of the LHC, the forward experiments FASER$\nu$ and SND@LHC will be able to detect the Charged Current (CC) interactions of the high energy neutrinos of all three flavors produced at the ATLAS Interaction Point (IP). This opportunity may unravel  mysteries of the third generation leptons. We build three models that can lead to a tau excess at these detectors through the following Lepton Flavor Violating (LFV) beyond Standard Model (SM) processes: (1) $\pi^+ \to \mu^+ \nu_\tau$; (2) 
	$\pi^+ \to \mu^+ \bar{\nu}_\tau$ and (3) $\nu_e+{\rm nucleus}\to \tau +X$. We comment on the possibility of solving the  $(g-2)_\mu$ anomaly and the $\tau$ decay anomalies within these models. We study the potential of the forward experiments  to discover the $\tau$ excess or to constrain these models in case of  no excess. We then  compare  the reach of the forward experiments with that of the previous as well as next generation experiments such as  DUNE. We also discuss how the upgrade of FASER$\nu$ can distinguish between these models by studying the energy spectrum of the tau. 
\end{abstract}

\maketitle


\section{Introduction}

Among the three neutrinos in nature, the tau-neutrino is the least studied one. Although the existence of $\nu_\tau$ had been established by the precise measurement of the $Z$ boson invisible decay width, its direct detection ({\it i.e.,} detection of $\tau$ from the Charged Current (CC) interaction of $\nu_\tau$) was announced  only in the early 21st century by the DONUT experiment at FermiLAB \cite{Kodama:2000mp}. Indeed, the $\tau$ data sample does not still exceed $\sim 21$ events, consisting of the  9 DONUT events \cite{Kodama:2007aa}, the 10 $\nu_\tau$ events registered by OPERA long baseline experiment \cite{Agafonova:2018auq} and two candidate events by ICECUBE \cite{Abbasi:2020zmr}.
The main reason why registering $\nu_\tau$ events is so difficult is that the produced $\tau$ at low energies is too short-lived to lead to a discernible track. Moreover, the conventional sources for neutrinos such as nuclear beta processes, muon decay or pion and Kaon decay produce only neutrinos of the first or second generations. The $\nu_\tau$ detected by OPERA comes from the oscillation of $\nu_\mu$ produced at CERN SPS en route to the detector at the Gran Sasso underground lab in Italy. 


The FASER$\nu$ \cite{Abreu:2019yak} and SND@LHC \cite{SND,Kling:2021gos} detectors during the run III of the LHC (2022-2024) will bring about a breakthrough in studying $\nu_\tau$. FASER$\nu$ and SND@LHC   are dense detectors,  designed to detect (and distinguish) all three kinds of neutrinos.  These experiments can also be sensitive to a variety of new physics involving dark matter \cite{Bakhti:2020vfq,Batell:2021blf,Ismail:2020yqc,Mitsou:2020okk,Arguelles:2019xgp} or beyond SM interaction of $\nu_\mu$ \cite{Ansarifard:2021elw,Bakhti:2020szu,Kling:2020iar, Falkowski:2021bkq,Kling:2021gos,Jodlowski:2020vhr,Bahraminasr:2020ssz,Beni:2020yfy}.

In this paper, we explore three new scenarios that can lead to the overproduction of the $\tau$ events at forward experiments, FASER$\nu$ and SND@LHC. We build models for these scenarios based on adding new scalar doublets to the SM. We show how by imposing global $U(1)$ flavor symmetries, the desired
flavor structure of the Yukawa coupling can be obtained.
As a bonus, these symmetries can explain the smallness of the first generation leptons and quarks.
 In each case, we show that how present experimental and observational constraints can be avoided and suggest strategies to test the accompanying prediction of the model by various experiments. 

The scenarios are the following: (1) $\pi^+\to \mu^+\nu_\tau$ with a branching ratio of $\sim 10^{-3}$. We show that this process can be obtained by adding  scalar doublets to the SM such that their charged components are mixed. Despite the stringent bounds from the processes such as $\tau^+\to \mu^+ \pi^0$,  we show that within our model $Br(\pi^+\to \mu^+\nu_\tau)\sim 10^{-3}$ can be achieved. (2)  $\pi^+\to \mu^+\bar{\nu}_\tau$ with again $Br(\pi^+\to \mu^+\bar{\nu}_\tau)\sim 10^{-3}$. The model that we build to embed this scenario involves a singlet charged scalar with an  asymmetric coupling to the second and third generation of left-handed leptons. 
Such a coupling has been proposed in \cite{Crivellin:2020klg} to explain the anomalies observed in the tau decay.
 (3) $\tau$ production via $\nu_e$ ($\nu_\mu$) scattering off the matter fields. In the model that we build for this scenario, $\tau$ and $\nu_e$ ($\nu_\mu$) have a Yukawa coupling with a new scalar doublet. We discuss the present bounds from the NOMAD data on the cross section of this process and then derive improvements that can be brought about by the upcoming FASER$\nu$ and SND@LHC experiments.
 
Ref. \cite{Falkowski:2021bkq}  discusses the bounds to be derived  from FASER$\nu$ on the effective couplings that can lead to processes $\pi^+ \to \mu^+ \nu_\tau$ and $\nu_e+{\rm nucleus}\to \tau+X$. The bounds that we have found for FASER$\nu$ are in  good agreement with theirs. We proceed with deriving the shape of the spectrum of $\tau$ for each scenario and comparing with the background $\tau$ spectrum within the Standard Model. We show that studying the spectrum during the high luminosity phase of the LHC at FASER$\nu$ 2 will dramatically increase the sensitivity to 
new physics. We also discuss the impact of the uncertainty in the prediction of the $\nu_\tau$ flux within the Standard Model.

 We  show that the effects of $\pi^+ \to \mu^+ \bar{\nu}_\tau$ and the $\tau$ production by $\nu_e$ can be also described  in terms of Charged Current (CC) Non-Standard Interaction (NSI) and the modified coherent source and detector eigenstates. In other words, we build  viable models  for sizable CC-NSI with observable effects at long baseline experiments such as DUNE.  
 
 This paper is organized as follows. In sections \ref{muNUtau}, \ref{NUbar} and \ref{TauProd}, we describe the models that give rise to the $\tau$ excess at forward experiments as mentioned above. We outline the parameter ranges that lead to a sizable excess and discuss their predictions for the CMS and ATLAS, anomalous muon magnetic dipole  moment and rare decays of the tau. In sect.~\ref{CC-NSI}, we show how the effects predicted by these models can be described within the well-studied formalism of  the Charged Current (CC) Non-Standard Interaction (NSI). We show that, thanks to a $m_\pi^2/(m_u+m_d)^2$ enhancement, we can obtain sizable CC NSI. We interpret the constraints on the CC NSI as bounds on our model. In sect.~\ref{spectra}, we derive the spectrum of $\tau$ within each model and discuss how the spectrum can help to discriminate between the  Standard Model (SM) background for the $\tau$ events   and the signals. 
 In sect.~\ref{FP}, we describe the relevant characteristics of the forward experiments of our interest and show that during the run III, FASER$\nu$ can significantly  reduce the uncertainty in the SM prediction for the number of the $\tau$ events.
 In sect.~\ref{signature}, we discuss the signature of the models in the forward experiments and present our results for the upcoming SND@LHC and  FASER$\nu$ experiments as well as for the  FASER$\nu$ upgrade  with higher statistics.  A summary and discussion is given in sect.~\ref{summary}.
 \section{The model(s) \label{model}}
 
 In this section, we introduce the models for (i) $\pi^+\to \mu^+ \nu_\tau$; (ii) $\pi^+\to \mu^+ \bar{\nu}_\tau$  and (iii) the $\tau$ production by $\nu_e$ ($\nu_\mu$)  scattering off the matter fields. In each case, we review the bounds on the parameter space of the model.
 We then show how the effects of these new models in the neutrino experiments can be described by the coherent $|\epsilon^d\rangle$ and $|\epsilon^s\rangle$ states that have  extensively been used in the literature to describe the CC-NSI.
 
 {\subsection{A model for $\pi^+ \to \mu^+ \nu_\tau$ \label{muNUtau}}
 	 The $\pi^+ \to \mu^+ \nu_\tau$ process is constrained by the precision measurement of the ratio $Br(\pi\to e \nu)/Br(\pi\to \mu \nu)$ where $\nu$ can be  any neutral fermion with a mass below 1 MeV that appears as missing energy. Notice that the SM prediction for this ratio is free from the uncertainties in the pion decay constant. The measurement is compatible with the SM prediction to the level of $2.4\times 10^{-3}$ \cite{Aguilar-Arevalo:2015cdf}, implying that $Br(\pi^+ \to e^+ \nu_\tau)<2.4\times 10^{-3}Br(\pi^+ \to e^+ \nu_e)=2.8\times 10^{-7}$ and  $Br(\pi^+ \to \mu^+ \nu_\tau)<2.4\times10^{-3}Br(\pi^+ \to \mu^+ \nu_\mu)=2.4\times10^{-3}$. \footnote{In this conclusion, we dismiss the accidental possibility that $Br(\pi^+ \to e^+ \nu_\tau)/Br(\pi^+ \to e^+ \nu_e)=
 Br(\pi^+ \to \mu^+ \nu_\tau)/Br(\pi^+ \to \mu^+ \nu_\mu)$. If this equality holds, the constraint on $Br(\pi\to e \nu)/Br(\pi\to \mu \nu)$ does not constrain  $Br(\pi^+ \to \mu^+ \nu_\tau)$ or $  Br(\pi^+ \to e^+ \nu_\tau)$, separately.  } Since the bound on $Br(\pi^+ \to e^+ \nu_\tau)$ is too strong to lead to an observable effect at FASER$\nu$ and other similar experiments, we will only focus on $\pi^+ \to \mu^+\nu_\tau$. 

The effective four-Fermi coupling 
\be G_{\nu \mu}(\bar{\mu}\frac{1-\gamma_5}{2}\nu_\tau )(\bar{d} \frac{1\pm \gamma_5}{2} u)\label{eff-pi-mu}\ee
leads to
\be \Gamma (\pi^+ \to \mu^+ \nu_\tau)=G_{\nu \mu}^2 \frac{m_\pi}{32 \pi}
\frac{F_\pi^2}{(m_u+m_d)^2} (m_\pi^2-m_\mu^2)^2. \label{PitomuNuTAU}\ee
With $ G_{\nu \mu} \sim 4\times 10^{-8}~{\rm GeV}^{-2}$, $Br (\pi^+ \to \mu^+ \nu_\tau)\sim 10^{-3}$.
Notice that although the $G_{\nu \mu}$ coupling  is chirality-flipping, the angular momentum conservation and the fact that both interactions are short-ranged imply that the polarizations of the muons emitted in $\pi^+ \to \mu^+ \nu_\tau$ and $\pi^+ \to \mu^+ \nu_\mu$ are equal. As a result, the precise measurement of the muon polarization \cite{pdg} does not constrain $G_{\nu \mu}$. 

To obtain the effective coupling in Eq (\ref{eff-pi-mu}), we introduce two scalar doublets, $\Phi_1=(\phi_1^+ \ \phi_1^0)^T$ and $\Phi_2=(\phi_2^+ \ \phi_2^0)^T$ with the following Yukawa couplings with the doublets, $L_\tau=(\nu_\tau \ \tau_L)^T$ and $Q_1=(u_L \ d_L)^T$:
\be \label{main-Yu} \lambda_d \bar{d} \Phi_1^\dagger Q_1+\lambda_u \bar{u} \Phi_1^T c Q_1+\lambda_\mu \bar{\mu}\Phi_2^\dagger L_\tau +{\rm H.c.} ,\ee
 where $c$ is an asymmetric matrix with $c_{12}=-c_{21}=1$.
 If $\Phi_1$ is identified with $\Phi_2$ or if the neutral components of these two doublets are mixed, the effective LFV $ G_\pi (\bar{\mu}_R \tau_L)(\bar{u}\gamma_5 u -\bar{d}\gamma_5 d)$  and
 ${G_\eta} (\bar{\mu}_R \tau_L)(\bar{u}\gamma_5 u +\bar{d}\gamma_5 d)$ couplings can be obtained by integrating out the heavy states.
  $G_\eta$ and $G_\pi$ will be respectively proportional to $\lambda_u+\lambda_d$ and $\lambda_u-\lambda_d$. These effective couplings lead to $\tau \to \mu \pi^0$ and $\tau \to \mu \eta^0$
 which are severely constrained \cite{pdg} and set bounds:
 $ G_\pi < 5 \times 10^{-9} ~~{\rm GeV}^{-2} $ and $ G_\eta <4\times 10^{-10}~~{\rm GeV}^{-2}.$
 To obtain $Br(\pi^+ \to \mu^+ \nu_\tau)\sim 10^{-3}$, we therefore need $G_{\nu \mu}\gg G_\pi , G_\eta$. This in turn implies $\Phi_1 \ne \Phi_2$. Moreover, the mixing between the neutral components of $\Phi_1$ and $\Phi_2$ should be much smaller than that between their charged components.\footnote{Notice that the mixing between the charged components can lead to a mixing between the neutral components  at one loop, suppressed by $e^2 \sin^2\theta_W/16\pi^2\sim 10^{-2}$ which is small enough.}
 
 To explain the flavor structure of the Yukawa couplings and to simplify the Lagrangian by removing unwanted terms, we impose an approximate $U_1(1)\times U_2(1)$ global symmetry. The $U_1(1)\times U_2(1)$ charges of the relevant fields are shown in table \ref{tab:bench}. The rest of the  SM fields  are neutral under this new $U_1(1)\times U_2(1)$.
 \begin{table}[h!]
 	\caption{The $U_1(1)\times U_2(1)$ charges of the fields. The rest of the fields, including $u_R$ and the Higgs are taken neutral under $U_1(1)\times U_2(1)$. }
 	\begin{center}
 	{\begin{tabular}{|c||*{6}{c|}}\hline
 				\centering
 				{\small charges}
 				&{$\Phi_1$}&{$\Phi_2$}
 				&{$L_\tau, \tau_R$}
 				&{$L_\mu,\mu_R$}&{$Q$}&
 				{$d_R$} \\\hline\hline
 				$U_1(1)$ & 1 & 0 & 0 & 0 & $\beta$ & $\beta-1$  \\\hline
 				$U_2(1)$ & 0 & 1 & $\alpha$ & $1+\alpha$ & 0 & 0  \\\hline
 				
 			\end{tabular}}
 		\end{center}
 		\label{tab:bench}
 	\end{table}%
 With this assignment, $\lambda_d\ne 0$ but $\lambda_u=0$ so our analysis will be simplified. Notice that the Yukawa couplings of $u$ and $d$ to the SM Higgs breaks the $U_1(1)$ symmetry so the smallness of the $u$ and $d$ masses can be explained as a bonus in this model. We can proceed with assigning unequal $U_1(1)\times U_2(1)$ charges to $e_R$ and $L_e$ to also explain the lightness of the first  generation of leptons but this is not the main goal of the present paper. The $U_1(1)\times U_2(1)$ symmetry explains the flavor structure of the Yukawa couplings and forbids mixing terms between $\Phi_1$ and $\Phi_2$ such as $\Phi_1^\dagger \Phi_2$,
 $|H|^2\Phi_1^\dagger \Phi_2$ and $(H^\dagger \Phi_1)(\Phi_2^\dagger H)$. As a result, $\phi_1^0$ and $\phi_2^0$ will not be mixed, preventing $\tau \to \mu \pi^0$.
 
 The mixing between $\phi_1^+$ and $\phi_2^+$, which is required to obtain $\pi^+\to\mu^+ \nu_\tau$, breaks $U_1(1)\times U_2(1)$.
 After electroweak symmetry breaking, we can obtain such a mixing between the charged components of $\Phi_1$ and $\Phi_2$ without mixing their neutral components  via the following term
 \be \lambda_{12}(H^T c\Phi_1)(\Phi_2^\dagger cH^*) \label{L12}.\ee
 Notice that this term explicitly breaks the global $U_1(1)\times U_2(1)$ to a single $U(1)$ under which $\Phi_1$ and $\Phi_2$ have equal charges. The effective coupling $G_{\nu \mu}$ can be written as
 \be  G_{\nu \mu}=\frac{\lambda_\mu\lambda_d}{m_{\phi_1^+}^2}
 \frac{\lambda_{12}v^2/2}{m_{\phi_2^+}^2}=4 \times 10^{-8} ~{\rm GeV}^{-2}\frac{\lambda_\mu}{0.3} \frac{\lambda_d}{0.3} \frac{\lambda_{12}}{0.12}\frac{(300~{\rm GeV})^2}{m_{\phi_1^+}^2}\frac{(300~{\rm GeV})^2}{m_{\phi_2^+}^2}. \label{Gnumu}\ee
 
 The $\lambda_\mu$ coupling can also give a contribution to $(g-2)_\mu$ of $\Delta a_\mu \sim \lambda_\mu^2 m_\mu^2/(100 \pi^2 m_{\Phi_2}^2)$ \cite{g-2-th}. In order to account for the $(g-2)_\mu$ anomaly \cite{g-2-exp} with $m_{\Phi_2}\sim 300$ GeV, $\lambda_\mu$ should saturate the perturbativity bound:
  $\lambda_\mu \sim 3$. In fact, this is a general feature of the models that explain the $(g-2)_\mu$ anomaly with new Yukawa coupling \cite{Allwicher:2021rtd}. To maintain $G_{\nu \mu}\sim 4 \times 10^{-8}$ GeV$^{-2}$, we can decrease $\lambda_{12}$ by one order of magnitude. The smallness of $\lambda_{12}$ can be explained by $U_1(1)\times U_2(1)\to U(1)$. 
 
 The components of  $\Phi_2$ can be pair produced at the LHC via the electroweak interactions. They will subsequently decay as $\phi_2^0 \to \mu^+\tau^-$ and 
 $\phi_2^+ \to \mu^+\nu_\tau$. The components of $\Phi_1$ can also be pair produced via the electroweak interactions. Moreover, the $\bar{d}+u$ and $\bar{d}+d$ scatterings can respectively produce $\phi_1^+$ and $\phi_1^0$ in association with the gluon.
 The $\Phi_1$ components will subsequently decay into a pair of jets.
 Through the mixing between $\Phi_1$ and $\Phi_2$, the electroweak interaction can also produce $\Phi_1 \Phi_2$ pairs. Moreover, the mixing can lead to the leptonic (hadronic) decay modes for $\Phi_1$ ($\Phi_2$). These effects are however subdominant and further suppressed by $O[(\lambda_{12} v^2/m_{\Phi_{1,2}}^2)^2]$. The heavier component of $\Phi_1$ or $\Phi_2$ can also decay into the lighter one and the $W$ boson. The splittings between the two components are however constrained by the oblique parameters 
 \cite{Haller:2018nnx}.
 The signature of pair production of the $\Phi_1$ as 
 well as single $\Phi_1$ production in association of gluon(s) will be multijet signal which suffers from high background. To our best knowledge, $\phi_1$ heavier than 200 GeV decaying into jets is still unconstrained by the LHC.  However, it may be discovered during the high luminosity phase of the LHC.  
 The signatures of the  $\phi_2^+ (\phi_2^0)^\dagger$, $ \phi_2^-\phi_2^0$, $ \phi_2^+\phi_2^-$ and $(\phi_2^0)^\dagger\phi_2^0$ are respectively  $\mu^+ \nu_\tau \mu^- \tau^+$, $\mu^- \bar{\nu}_\tau \mu^+ \tau^-$, $\mu^+ \nu_\tau \mu^- \bar{\nu}_\tau$ and $\mu^- \tau^+ \mu^+ \tau^-$ where the invariant masses of the $\tau $ and $\mu$ pairs are equal to $m_{\phi_2^0}$. To our best knowledge, neither a  dedicated search for $\phi_2^0$  with an arbitrary mass decaying into $\mu^+\tau^-$ nor a search for $\phi_2^+$ decaying into the muon plus missing energy has been carried out, yet. \footnote{ There is already a stringent bound on  the LFV decay mode  of the SM Higgs:  $Br(H\to \tau \mu)< 0.28 \%$ \cite{Aad:2019ugc}. This bound  can be translated into an upper bound on the mixing between $H^0$ and $\phi_1^0$. Such a mixing   violates the global $U_1(1)\times U_2(1)$ symmetry as well as the residue $U(1)$ that survives the introduction of $\lambda_{12}$. Thus, in our model, the mixing between $H^0$ and $\phi_1^0$ is naturally small.}   
 
 Notice that $\pi^+ \to \mu^+ \nu_\tau$ is enhanced by $f_\pi^2/(m_u +m_d)^2$ but the cross section of the $\nu_\tau$ interaction on the nuclei via the new $G_{\nu \mu}$ does not enjoy such as enhancement. Moreover, since there is a large background for the ($\mu$+jets) signal from the CC interaction of $\nu_\mu$, we do not need to worry about the impact of $G_{\nu \mu}$ on the detection.

In sect. \ref{signature}, we shall study the bounds from FASER$\nu$  on $Br(\pi^+ \to \mu^+ \nu_\tau)$.
This scenario could lead to the tau production at the NOMAD detector, too. However, at NOMAD the energies of neutrinos from the pion decay are around 20~GeV so the momentum of the jets recoiling against the produced $\tau$ would be too low to survive the cuts applied by the NOMAD collaboration to identify the $\tau$ production \cite{Astier:2001yj}. At NOMAD, the neutrino flux with energies higher than 50 GeV was also produced but the production was dominated by the Kaon decay rather than the pion decay. As a result, the exotic decay $K^+ \to \mu^+ \nu_\tau$ can already strongly  be constrained by NOMAD. We have therefore  focused only on  
	the exotic pion decay in this paper.	
\subsection{ A model for $\pi^+ \to \bar{\nu}_\tau \mu^+$ with a connection to observed anomalies in $\tau$ decay\label{NUbar}}
Ref. \cite{Crivellin:2020klg} proposes a model to address  the $2 \sigma$ discrepancy between the observation and the SM prediction in the 
$\tau \to \mu \nu_\tau \bar{\nu}_\mu$ mode \cite{Amhis:2019ckw}.
The model is based on the introduction of a new charged singlet $\Phi^+$ heavier than 300~GeV and with an interaction of form
\be \mathcal{L}=-\frac{\lambda_{23}}{2} L_{a,\mu}\epsilon_{ab} L_{b \tau} \Phi^++{\rm H.c}= -\frac{\lambda_{23}}{2} (\nu_\mu^T c\tau_L-\mu_L^T c \nu_\tau)\Phi^+ +{\rm H.c}\ee 
From  ${\rm Br}(\tau \to \mu \nu \nu)/{\rm Br}(\tau(\mu)\to e \nu \nu)$, Ref. 
\cite{Crivellin:2020klg} finds 
\be \label{Lam23} 0.052 \frac{m_{\Phi^+}}{300~{\rm GeV}}
<\lambda_{23}<0.148 \frac{m_{\Phi^+}}{300~{\rm GeV}}. \ee
 The $\lambda_{23}$ coupling can also give rise to $(g-2)_\mu$ but considering the upper bound on $\lambda_{23}$ shown in Eq.
 (\ref{Lam23}), the contribution will be too small to account for the observed deviation from the standard model prediction \cite{g-2-th,g-2-exp}.
 
 \begin{table}[h!]
 	\caption{The $U(1)$ charges of the fields. The rest of the fields, including $u_R$ and the Higgs are  neutral under $U(1)$. }
 	\begin{center}
 		{\begin{tabular}{|c||*{6}{c|}}\hline
 				\centering
 				{\small charges}
 				&{$\Phi_1$}&{$\Phi^+$}
 				&{$L_\tau, \tau_R$}
 				&{$L_\mu,\mu_R$}&{$Q$}&
 				{$d_R$} \\ \hline\hline
 				$U(1)$ & 1 & 1 & $-1/2-\alpha$ & $-1/2+\alpha$ & $\beta$ & $\beta-1$  
 				\\\hline
 			\end{tabular}}
 		\end{center}
 		\label{tab:SingleU(1)}
 	\end{table}%

 	Let us reintroduce $\Phi_1$ of section \ref{muNUtau} to this section with the $U(1)$ charges as in Table  \ref{tab:SingleU(1)}. With this assignment, we can have a trilinear term as 
 	$$ A\Phi^- H^Tc\Phi_1$$ which after electroweak symmetry breaking induces a mixing between $\Phi^+$ and $\phi_1^+$ given by
 	\be \sin 2 \theta=\frac{2A~ v/\sqrt{2}}{m_{\phi_1^+}^2-m_{\Phi^+}^2}. \ee
 	Integrating out the heavy fields, we shall have an effective coupling of form
\be G_{\bar{\nu}\mu} (\bar{d}\frac{1-\gamma_5}{2}u)(\nu_\mu^T c\tau_L-\mu_L^T c \nu_\tau)+{\rm H.c.}\label{eff-pi-bar}\ee
where $$G_{\bar{\nu}\mu}=\frac{\lambda_d\lambda_{23} }{2}\frac{A v/\sqrt{2}}{m_{\Phi^+}^2m_{\phi_1^+}^2 }.$$
With this effective Lagrangian, a new decay mode $\pi^+ \to \bar{\nu}_\tau \mu^+$ will open with a rate given by Eq. (\ref{PitomuNuTAU})  but replacing $G_{\nu \mu}$ with $G_{\bar{\nu}\mu}$. Similarly to the decay via $G_{\nu \mu}$, with $G_{\bar{\nu}\mu}\sim 5 \times 10^{-8}$ GeV$^{-2}$, Br$(\pi^+ \to \bar{\nu}_\tau \mu^+)$ can be as large as $10^{-3}$. 

The axial component of  $G_{\bar{\nu}\mu}$ can  lead to \be \label{TauTOpiII} \Gamma(\tau^- \to \bar{\nu}_\mu \pi^-)\sim \frac{G_{\bar{\nu}\mu}^2}{4\pi}\frac{F_\pi^2 m_\pi^2}{(m_u+m_d)^2} m_\tau \ee  which is again enhanced by $ m_\pi^2/(m_u+m_d)^2$. The corresponding  branching ratio is  ${\rm Br}(\tau^- \to \bar{\nu}_\mu \pi^-)\sim 5 \times 10^{-6} [G_{\bar{\nu}\mu}/(5 \times 10^{-8}~ {\rm GeV}^{-2})]^2$ which is much smaller than the uncertainty in ${\rm Br}(\tau \to \pi \nu)$ which is $5\times 10^{-4}$ \cite{pdg}. 
Within the SM, the branching ratio of $\tau^-\to \nu_\tau \pi^0\pi^-$ is even larger than that of the two body decay  $\tau^-\to \nu_\tau \pi^-$. The enhancement is  due to the spin 1 $\rho$ resonance from the vectorial part of the charged current, $\tau^- \to \nu_\tau \rho^-\to \nu_\tau \pi^0\pi^-$  \cite{Burchat:1986na,Braguta:2004kx}. In our model, since the mediator ($\Phi^+$) has zero spin, no $\rho$ resonance occurs so we expect $\tau^- \to \bar{\nu}_\mu \pi^0\pi^-$ to be suppressed. 
 Via the $G_{\bar{\nu}\mu}$ interaction, $\nu_\mu$ can produce $\tau$ in the detector, too, but the cross section will be suppressed by $\sim G_{\bar{\nu}\mu}^2/(8G_F^2)\sim  2\times 10^{-6}$  relative to the SM CC interaction of $\nu_\mu$. This means the number of $\tau$ events produced during the run III of the LHC by the $\nu_\mu$ flux  will be as small as $O(0.01)$ and therefore negligible.
Similarly, the bound on the $\tau$ production at NOMAD \cite{Astier:2001yj} can be avoided.

In this model, $\Phi^+$ and $\Phi^-$ can be pair produced at the HL-LHC by electromagnetic interactions. They will then decay as $\Phi^+ \to \mu^+ \nu$ and $\Phi^+ \to \tau^+ \nu$ so the signals will be excess in the $\mu^+\mu^-+{\rm missing ~energy}$,
$\tau^+\tau^-+{\rm missing ~energy}$, $\tau^- \mu^++{\rm missing ~energy}$ and $\tau^+\mu^-+{\rm missing ~energy}$ signals. The  $\Phi_1$ pairs can also be produced at the LHC, decaying into jets as described in the previous subsection.

{\subsection{Non-standard $\tau$  production at the detector \label{TauProd}}

In this section, we introduce a variation of the model introduced in sect. \ref{muNUtau} with the difference that $\Phi_2$ couples to $\tau_R$ instead of $\mu_R$ as follows
\be \lambda_e \bar{\tau}_R \Phi_2^\dagger L_e+  \lambda_\mu \bar{\tau}_R \Phi_2^\dagger L_\mu
.\ee If  $\lambda_e$ and $\lambda_\mu$ are both nonzero, they can contribute to $\mu \to e \gamma$ at one loop which is severely constrained by the experimental bounds. As a result, we assume that only one of $\lambda_e$ and $\lambda_\mu$ is nonzero. This pattern
can be explained by the $U_2(1)$ symmetry.
 For example, if we assign 
  $U_2(1)$ charges  to $\Phi_2$ and leptons as shown in Table \ref{third}, we can simultaneously  explain nonzero $\lambda_e$, vanishing $\lambda_\mu$
  and the smallness of the electron mass.
   \begin{table}[h!]
   	\caption{The $U_2(1)$ charges of the fields.  The rest of the fields, including the second generation leptons, $\Phi_1$, quarks and the Higgs are  neutral under $U_2(1)$. }
   	\begin{center}
   		{\begin{tabular}{|c||*{6}{c|}}\hline
   				\centering
   				{\small charges}
   				&{$\Phi_2$}
   				&{$L_\tau, \tau_R$}
   				&{$L_e$}&{$e_R$, $H$, quarks, $L_\mu$, $\mu_R$, $\Phi_1$} \\ \hline\hline
   				$U_2(1)$&1 & b & 1+b &0  
   				\\\hline
   			\end{tabular}}
   		\end{center}
   		\label{third}
   	\end{table}%
  
   Like the model in sect. \ref{muNUtau}, we allow only the charged components to mix with each other. As a result, the severely constrained decay modes $\tau^- \to e^- \pi^0$ or $\tau^- \to \mu^- \pi^0$ cannot be obtained at the tree level. However, we obtain \be 
 \label{GeGmu}G_e (\bar{\tau}_R \nu_e)( \bar{u}_L d_R) \ \ \ {\rm or }    \ \ \ G_\mu (\bar{\tau}_R \nu_\mu) (\bar{u}_L d_R)\ee  where $ G_{e}=\lambda_d \lambda_e \lambda_{12}v^2/(2m_{\phi_1^+}^2m_{\phi_2^+}^2)$ and $ G_{\mu}=\lambda_d \lambda_\mu \lambda_{12}v^2/(2m_{\phi_1^+}^2m_{\phi_2^+}^2)$. These effective couplings respectively lead to $\tau^- \to \pi^- +\nu_e$ and $\tau^- \to \pi^- +\nu_\mu$. Similarly to sect. \ref{muNUtau} and the case of $G_{\bar{\nu}\mu}$ in Eq. (\ref{TauTOpiII}), the uncertainty on $\tau^+\to \pi^+ \nu$ gives the constraint $G_{e (\mu)}< 5 \times 10^{-7}$ GeV$^{-2}$.\footnote{Notice that the bound that we have found on $G_e$ from $\tau^+ \to \pi^+ \nu_e$ is much stronger than the bound in \cite{Falkowski:2021bkq}. To derive this bound we have equated $Br(\tau^+ \to \pi^+\nu_e$) with the experimental uncertainty in $Br(\tau^+\to \pi^+ +\nu)$ which is $5\times 10^{-4}$.} Saturating this constraint, we shall have $\sigma(\nu_{e(\mu)}+{\rm nucleus}\to \tau +X)/\sigma(\nu_\mu+{\rm nucleus}\to \mu +X)\sim (G_{e(\mu)}/4G_F)^2\sim  10^{-4}$. In this model, regardless of the origin of the neutrinos (whether they come from the pion or Kaon decays), the electron or muon neutrinos with energies sufficiently  larger than the tau mass can lead to the production of $\tau$. As a result, the NOMAD experiment can constrain $G_e$ and $G_\mu$ (cf. the model in sect 
 \ref{muNUtau} which  avoids the NOMAD constraints as explained.)
 The number of the $\nu_\mu$ charged current events with an energy larger than  25 GeV  observed at NOMAD was  above $2\times 10^5$  which
  is one order of magnitude larger than the anticipated number at FASER$\nu$ during run III. The bound from NOMAD on $G_\mu$ would therefore be of order of $5\times 10^{-8}$ GeV$^{-2}$ which is even stronger than the bound from $\tau^+ \to \pi^+ \nu_\mu$. Such a strong bound on $G_\mu$  makes observing a deviation from the SM prediction at FASER$\nu$ hopeless so we shall not study the effects of $G_\mu$ at FASER$\nu$ any further. On the other hand, the number of the $\nu_e$ events
 at NOMAD and FASER$\nu$ are comparable so the bound on $G_e$ may be improved by FASER$\nu$. In sect \ref{CC-NSI}, we will quantify  the bound from NOMAD on $G_e$. We shall study the bound that FASER$\nu$ and its upgrades can set on 
 $\sigma (\nu_e +{\rm nucleus}\to \tau +X)$ in sect. \ref{summary}.

 \subsection{Connection to the Charged Current Non-Standard Interaction formalism\label{CC-NSI}}
 
 There is a rich literature studying the Non-Standard Interaction (NSI)
 on neutrino oscillation experiments \cite{Farzan:2017xzy}. The effects of Charged Current NSI are often analyzed by introducing eigenstates of source and detector as follows
 \be |\nu_\alpha^s\rangle=|\nu_\alpha \rangle+\sum_{\gamma \in \{e,\mu,\tau \}}\epsilon^s_{\alpha \gamma}
 |\nu_\gamma\rangle \label{nuAs}\ee
 and 
 \be \langle \nu_\alpha^d|=\langle\nu_\alpha |+\sum_{\gamma \in \{e,\mu,\tau \}}\epsilon^d_{\gamma \alpha}
 \langle\nu_\gamma| \label{nuAd}\ee where $|\nu_\alpha^s\rangle$ is the eigenstate produced in the source along with the charged lepton of flavor $\alpha$ and $|\nu_\alpha^d\rangle$  is the eigenstate which can produce the charged lepton of flavor $\alpha$ in the detector. Within the SM, $|\nu_\alpha^s\rangle=|\nu_\alpha^d\rangle=|\nu_\alpha\rangle$. However, non-standard interaction can in principle induce nonzero $\epsilon_{\alpha \beta}^s$ and $\epsilon_{\alpha \beta}^d$.  In recent years, a class of models have been developed based on a new light neutral $U(1)$ gauge boson coupled to neutrinos and matter fields that induces a sizable neutral current NSI \cite{khodam}. In case of CC NSI, the mediator has to be a charged particle so its mass must be heavier than a few 100 GeV to avoid direct production at the LEP and/or at the LHC.  Since the relevant effective four-Fermi coupling is given by inverse of the square of the mediator mass, a strong lower bound on the mediator mass generally means small CC NSI. With this consideration, not many models are proposed to underly the CC NSI, despite the extensive efforts to study their phenomenological impact on the neutrino experiments. Indeed, $G_{\nu \mu}$  obtained in Eq. (\ref{Gnumu}) is quite suppressed $G_{\nu \mu}\ll G_F$. Despite the smallness of $G_{\nu\mu}$, thanks to the  $m_\pi/(m_u+m_d)$ enhancement in the amplitude of $\pi^+\to \mu^+ \nu_\tau$ relative to that of the standard $\pi^+\to \mu^+ \nu_\mu$, ${\rm Br}(\pi^+\to \mu^+ \nu_\tau)$ can be still relatively large. Within the model introduced in sect.~\ref{muNUtau}, $|\nu_\mu^s\rangle$ can be written as 
 \be \label{NusMu} |\nu_\mu^s\rangle=\frac{\mathcal{M}_1|\nu_\mu\rangle+\mathcal{M}_2|\nu_\tau\rangle }{\sqrt{|\mathcal{M}_1|^2+|\mathcal{M}_2|^2}}\simeq |\nu_\mu\rangle+\mathcal{M}_2/\mathcal{M}_1|\nu_\tau\rangle,\ee
 where $\mathcal{M}_1$ and $ \mathcal{M}_2$ are respectively the amplitudes of $\pi^+\to \mu^+ \nu_\mu$ and $\pi^+\to \mu^+ \nu_\tau$. Thus, in the model introduced in sect. \ref{muNUtau}, 
 $$ \epsilon_{\mu \tau }^s=\frac{\mathcal{M}_2}{\mathcal{M}_1}.$$ We can therefore write  $|\epsilon_{\mu \tau}^s|^2={|\mathcal{M}_2|^2}/{|\mathcal{M}_1|^2}\simeq {\rm Br}(\pi^+ \to \mu^+ \nu_\tau)$. In this model, $ \epsilon_{\mu \tau }^d \ll 1$.
 If the baseline of the experiment is short such that 
 $\Delta m_{atm}^2 L/E_\nu \ll 1$,  the number of the $\mu$ events and  the excess of the $\tau$
 events in the detector will respectively be given by ${\rm Br}(\pi^+ \to \mu^+ \nu_\mu)\sigma_{SM}(\nu_\mu \to \mu)$ and ${\rm Br}(\pi^+ \to \mu^+ \nu_\tau)\sigma_{SM}(\nu_\tau \to \tau)$.  Thus, it is valid to analyze the FASER$\nu$  results as well as the DUNE near detector data in terms of ${\rm Br}(\pi^+ \to \mu^+ \nu_\tau)$ rather than studying the evolution of the coherent state in Eq.~(\ref{NusMu}). However, for the long baseline experiments, it is necessary to study the evolution of the full coherent state in Eq.  (\ref{NusMu}); otherwise, we will miss the effect of the interference terms given by  $2Re[U_{\mu i}U_{\mu i}^*U_{\mu j}^*U_{\tau j}(\epsilon_{\mu \tau}^s)^* e^{i(m_i^2-m_j^2)L/(2E_\nu)}]$  in case of the $\mu$ detection and 
 $2Re[U_{\tau i}U_{\mu i}^*U_{\tau j}^*U_{\tau j}(\epsilon_{\mu \tau}^s)^* e^{i(m_i^2-m_j^2)L/(2E_\nu)}]$ in case of the $\tau$ detection. Notice that both these interference terms are linear in  $\epsilon_{\mu \tau}^s$ and therefore dominate over the effect of $Br(\pi^+ \to \mu^+ \nu_\tau)=|\epsilon_{\mu \tau}^s|^2$.
 
In case of $\nu_{e (\mu)}+{\rm nucleus}\to \tau +X$ within   the model introduced in sect.~\ref{TauProd}, we should pay attention that the chirality of $\tau$ produced via the new coupling is opposite to that produced by $\nu_\tau$ in the SM. As a result, the interference term will be suppressed by $m_\tau/E_\nu$ and we cannot therefore simply equate $\epsilon_{e(\mu)\tau}^d$  with $\mathcal{M}(
\nu_{e (\mu)}+{\rm nucleus}\to \tau +X)/\mathcal{M}(\nu_{\tau}+{\rm nucleus}\to \tau +X)$.  In fact, the helicity of the final $\tau$ has to be considered, too. For short baseline experiment such as FASER$\nu$ or NOMAD for which $\Delta m_{atm}^2 L/(2 E_\nu)\ll 1$, such interference is not relevant and we can use the bounds on $\epsilon_{e(\mu)\tau}^d$ and on $[\sigma(
\nu_{e (\mu)}+{\rm nucleus}\to \tau +X)/\sigma(\nu_{\tau}+{\rm nucleus}\to \tau +X)]^{1/2}
\simeq G_{e(\mu)}/(\sqrt{96} G_F)$, interchangeably.   
 	As discussed in Sect \ref{TauProd}, the NOMAD experiment  can constrain this model.  From the NOMAD data, Ref. \cite{Biggio:2009nt} finds $\epsilon_{e \tau }^d<0.087$ which implies $\sigma(
 	\nu_{e (\mu)}+{\rm nucleus}\to \tau +X)/\sigma(\nu_{\tau}+{\rm nucleus}\to \tau +{\rm nucleus})<0.0075$ and $G_{e(\mu)}/G_F<0.85$ which is readily satisfied in the model described in sect. \ref{TauProd}.


 	The far detector of DUNE can also constrain $\epsilon^d$ and $\epsilon^s$ \cite{Blennow:2016etl}. To study the effects at far detector of DUNE, the coherent states $|\nu^s\rangle$ and 
 	$|\nu^d\rangle$ have to be used.
 	 We could also define a coherent state of $(\mathcal{M}_1|\nu_\mu\rangle +\mathcal{M}_2|\bar{\nu}_\tau\rangle)/\sqrt{|\mathcal{M}_1|^2+ 
 	 	|\mathcal{M}_2|^2}$ to describe the effects of the model in sect. \ref{NUbar} but since no interference between evolved $|\nu_\mu \rangle$ and $|\bar{\nu}_\tau\rangle$ takes place even for long baselines, there is no point in introducing such a coherent state.
 	 
 	\section{Spectrum of $\tau$ produced at forward experiments \label{spectra}}
 	In this section, we compute the spectrum of $\tau$ produced  via different types of interaction introduced in this paper and compare with the tau spectrum produced  via the standard CC electroweak interactions. 
	
 	The $G_e$ coupling defined in Eq. (\ref{GeGmu}) leads to
 	\be \langle \left|\mathcal{M}[\nu_e(E_\nu)+d(x)\to \tau^{-}(E_\tau) +u]\right|^2\rangle=\langle\left|\mathcal{M}[\nu_e(E_\nu)+\bar{u}(x)\to \tau^{-}(E_\tau) +\bar{d}]\right|^2\rangle= \ee $$\langle\left|\mathcal{M}[\bar{\nu}_e(E_\nu)+\bar{d}(x)\to \tau^{+}(E_\tau) +\bar{u}]\right|^2\rangle=\langle\left|\mathcal{M}[\bar{\nu}_e(E_\nu)+{u}(x)\to \tau^{+}(E_\tau) +{d}]\right|^2\rangle=$$ 	
 	$$2G_{e}^2(P_\tau\cdot P_\nu)(P_u \cdot P_d) =2G_{e}^2 x^2 m_N^2(E_\nu -E_\tau)^2, $$
 	where $m_\tau^2/(2xm_N)<E_\tau <E_\nu$.
 	Thus, the  differential cross sections of all these four processes   can be written as 
 	\be \frac{d  \sigma}{dE_\tau}=\frac{1}{16 \pi}\frac{1}{E_\nu s}|\mathcal{M}|^2=\frac{G_{e}^2}{16 \pi}x m_N \left(1- \frac{E_\tau}{E_\nu}\right)^2 ~~~~~\frac{m_\tau^2}{2xm_N }<E_\tau <E_\nu. \label{sigScalar} \ee

 	We can then write 
 	\begin{eqnarray} 
 		\sigma(\nu_e+{\rm nucleus}\to \tau +X)&=& \frac{G_e^2}{48 \pi}m_NE_\nu \int_{x_{min}}^1 x\left( 1-\frac{m_\tau^2}{2xm_NE_\nu}\right)^3[F_d(x,t)+F_{\bar{u}}(x,t)]dx\label{17}\\
 		\sigma(\bar{\nu}_e+{\rm nucleus}\to \bar{\tau} +X)&=& \frac{G_e^2}{48 \pi}m_NE_\nu \int_{x_{min}}^1 x\left( 1-\frac{m_\tau^2}{2xm_NE_\nu}\right)^3[F_{\bar{d},t }(x,t)+F_{u}(x,t)]dx \label{18}
 			\end{eqnarray}
 		where $F_q$ is the $q$-quark parton distribution function and $$x_{min}=\frac{m_\tau^2}{2m_N E_\nu} \ \ \ \ {\rm and } \ \ \ \ t=2xm_N(E_\tau-E_\nu).$$
 		
 	The spectrum of $\tau+\bar{\tau}$ produced by $\nu_e +{\rm nucleus}\to \tau+X$ and $\bar{\nu}_e +{\rm nucleus}\to \bar{\tau}+X$ can be written as 
 	\be S_{e}(E_\tau)=\frac{ \int_{E_{\tau}} \int_{x_{min}}^1 [F_{\nu_e}(E_{\nu})(F_d+F_{\bar{u}})+F_{\bar{\nu}_e}(E_\nu)(F_{\bar{d}}+
 		F_{{u}})]\frac{d \sigma}{dE_\tau} ~ d E_{\nu} dx}{\int   \int_{x_{min}}^1 \int_{m_\tau^2/(2xm_N)}^{E_\nu} [F_{\nu_e}(E_\nu)(F_d+F_{\bar{u}})+F_{\bar{\nu}_e}(E_\nu)
 		(F_{\bar{d}}+
 		F_{{u}})]\frac{d \sigma}{dE_\tau} dE_\tau~dx~ d E_{\nu} }, \label{Eq:Se}\ee
 where $d\sigma/dE_\tau$  is given in Eq. (\ref{sigScalar}). The parton distribution functions are functions of both $x$ and  the Mandelstam variable, $t$.
 	
 	For comparison the standard model cross sections are 
 	\be \frac{d \sigma(\nu_\tau +d \to \tau^- +u)}{dE_\tau}=
 	\frac{d \sigma(\bar{\nu}_\tau +\bar{d} \to \tau^+ +\bar{u})}{dE_\tau}
 	=\frac{2m_N xG_F^2}{ \pi}\frac{m_W^4}{[2(E_\nu-E_\tau)x m_N +m_W^2]^2},\label{sigSM1}\ee 
 	and 
 	\be \frac{d \sigma(\bar{\nu}_\tau +u \to \tau^+ +d)}{dE_\tau}=
 	\frac{d \sigma({\nu}_\tau +\bar{u} \to \tau^- +\bar{d})}{dE_\tau}
 	=\frac{2m_N xG_F^2}{ \pi}
 	\left( \frac{E_\tau}{E_\nu}\right)^2 \frac{m_W^4}{[2(E_\nu-E_\tau)x m_N +m_W^2]^2},\label{sigSM2}
 	\ee 
 	where $$\frac{m_\tau^2}{2xm_N}<E_\tau <E_\nu.$$
 	Notice that while in Eq. (\ref{sigScalar}), we have used the effective four-Fermi coupling, $G_e$, in Eqs. (\ref{sigSM1},\ref{sigSM2}), we have used the full propagator for $W$.  This is understandable as for $x\sim 0.1$,  $2xm_NE_\nu \stackrel{<}{\sim} m_W^2\ll m_{\phi^+_1}^2,  m_{\phi^+_2}^2$. In fact, we have found that neglecting $2(E_\nu-E_\tau)xm_N$ in the denominator induces an error of 3\% in the total number of events.

 	The total cross section of $\nu_\tau$ and $\bar{\nu}_\tau$ scattering off the nucleon within the SM can be written as
 	$$\frac{d\sigma^{SM}_\nu}{dE_\tau}= \int_{x_{min}}^1 [F_d(x,t)\frac{d\sigma(\nu_\tau +d \to \tau^- +u)}{dE_\tau}+F_{\bar{u}}(x,t)\frac{d\sigma(\nu_\tau +\bar{u} \to \tau^- +\bar{d})}{dE_\tau} ]d x $$
 	and
 	$$\frac{d\sigma^{SM}_{\bar{\nu}}}{dE_\tau}= \int_{x_{min}}^1 [F_{\bar{d}}(x,t)\frac{d\sigma(\bar{\nu}_\tau +\bar{d} \to \tau^+ +\bar{u})}{dE_\tau}+F_{{u}}(x,t)\frac{d\sigma(
 		\bar{\nu}_\tau +{u} \to \tau^+ +{d})}{dE_\tau} ]d x .$$
 	Finally we can write 
 	 	$$\sigma^{SM}_\nu= \int_{x_{min}}^1\int^{E_\nu}_{m_\tau^2/(2xm_N)} [F_d(x,t)\frac{d\sigma(\nu_\tau +d \to \tau^- +u)}{dE_\tau}+F_{\bar{u}}(x,t)\frac{d\sigma(\nu_\tau +\bar{u} \to \tau^- +\bar{d})}{dE_\tau} ]dE_\tau d x  $$
 	 	and
 	 	a similar formula for $\sigma^{SM}_{\bar{\nu}}$ replacing particles with antiparticles.

 		As discussed in section \ref{muNUtau}, the effective $G_{\nu \mu}$ coupling introduced in Eq. (\ref{eff-pi-mu})
 		can also lead to the  $\tau$ production via charged pion decay.
 	The signal from $\pi^+ \to \mu^+ \nu_\tau$ and $\pi^- \to \mu^- \bar{\nu}_\tau$  will have the following form
 	\be S_{\nu \mu} (E_\tau)\equiv \frac{\int_{E_\tau} 
 		[F^\pi_{\nu_\mu}(E_\nu)
 		\frac{d\sigma^{SM}_{{\nu}}}{dE_\tau}+F^\pi_{\bar{\nu}_\mu}(E_\nu)
 		\frac{d\sigma^{SM}_{\bar{\nu}} }{dE_\tau}]dE_\nu}{\int    [F^\pi_{\nu_\mu}(E_\nu)
 		\sigma^{SM}_{{\nu}}+F^\pi_{\bar{\nu}_\mu}(E_\nu)
 		\sigma^{SM}_{\bar{\nu}} ] dE_\nu },\label{S2}\ee
 where $F^\pi_{\nu_\mu} (E_\nu)$ and  $F^\pi_{\bar{\nu}_\mu} (E_\nu)$ are the spectra of neutrinos from the pion decay (rather than the whole flux from  pion and Kaon decay).
 
 Let us now discuss the spectrum of the tau produced by lepton number and lepton flavor violating pion decay mode caused by the effective coupling $G_{\bar{\nu}\mu}$ introduced in Eq. (\ref{eff-pi-bar}) of section \ref{TauProd}. 
 	The signal from $\pi^+ \to \mu^+ \bar{\nu}_\tau$ and $\pi^- \to \mu^- {\nu}_\tau$  will have a form given by Eq
 	(\ref{S2}), swapping $F_{\nu_\mu}$ and $F_{\bar{\nu}_\mu}$:
 		\be S_{\bar{\nu}\mu} (E_\tau)\equiv \frac{\int_{E_\tau} 
 			[F_{\bar{\nu}_\mu}^\pi(E_\nu)
 			\frac{d\sigma^{SM}_{{\nu}}}{dE_\tau}+F_{{\nu}_\mu}^\pi(E_\nu)
 			\frac{d\sigma^{SM}_{\bar{\nu}} }{dE_\tau}]dE_\nu}{\int    [F^\pi_{\bar{\nu}_\mu}(E_\nu)
 			\sigma^{SM}_{{\nu}}+F^\pi_{{\nu}_\mu}(E_\nu)
 			\sigma^{SM}_{\bar{\nu}} ] dE_\nu }.\label{S3}\ee
 		
 	
 	Finally the $\tau$ spectrum within the standard model will have the form 
 	\be B=\frac{\int_{E_\tau} 
 		[F_{\nu_\tau}(E_\nu)
 		\frac{d\sigma^{SM}_{{\nu}}}{dE_\tau}+
 		F_{\bar{\nu}_\tau}(E_\nu)
 		\frac{d\sigma^{SM}_{\bar{\nu}} }{dE_\tau}]dE_\nu}{\int    [F_{{\nu}_\tau}(E_\nu)
 		\sigma^{SM}_{{\nu}}+F_{\bar{\nu}_\tau}(E_\nu)
 		\sigma^{SM}_{\bar{\nu}} ] dE_\nu }. \label{Back}\ee

From Eqs. (\ref{17},\ref{18},\ref{sigSM1},\ref{sigSM2}), we observe that the cross sections of all the processes are suppressed by $x$ for small values of $x$.
As a result, the main contribution to the cross section comes from $x\sim {\rm few}\times 10^{-2}-1$. Thus, $Q^2=-t=2(E_\nu -E_\tau)xm_N\sim 100~{\rm GeV}^2$. 

\begin{figure}
	\centering
	\includegraphics[width=0.6\textwidth]{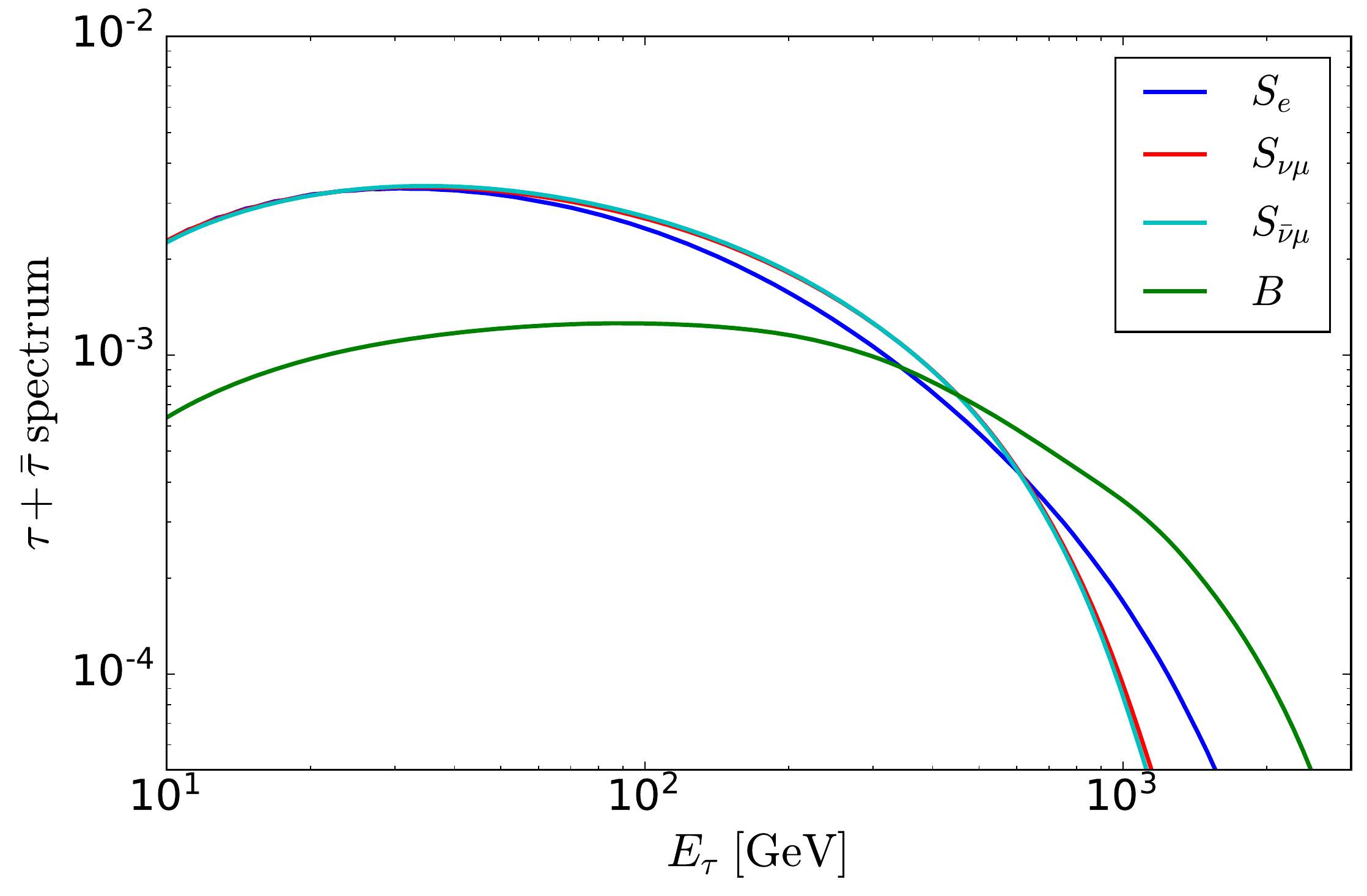}
	\caption{Spectra of $\tau+\bar{\tau}$ produced at FASER$\nu$ normalized to 1. The curves marked with $S_e$, $S_{\nu\mu}$ and
		$S_{\bar{\nu}\mu}$ show the spectra of $\tau+\bar{\tau}$ from new physics scenarios $\nu_e+{\rm nucleus}\to \tau+X$, $\pi^+\to \mu^+\nu_\tau$, $\pi^+\to \mu^+\bar{\nu}_\tau$, respectively. The standard model background is marked with $B$. The input neutrino  spectra that we insert  to draw the $\tau+\bar{\tau}$ spectrum ({\it i.e.,} $F_{\nu_\tau}$, $F_{\bar{\nu}_\tau}$  $F_{\nu_\mu}^\pi$ and $F_{\bar{\nu}_\mu}^\pi$) are  described in the last paragraph of sect.~\ref{FP}. }
	\label{fig:spectrum}
\end{figure}
 The normalized spectra of $\tau +\bar{\tau}$ from each scenario are shown in Fig.~\ref{fig:spectrum}. To draw the curves, we have averaged the scattering cross section over the protons and neutrons composing Tungsten nucleus. As seen from the figure, the background from SM is significantly harder than new physics. This is mostly due to the fact that the background comes from $F(\nu_\tau)$ and $F(\bar{\nu}_\tau)$ which are harder than the spectra of other neutrino flavors; {\it cf.,} Eq.~(\ref{Back})
with Eqs.~(\ref{Eq:Se},\ref{S2},\ref{S3}). The spectrum of background is quite distinct from $S_{\nu\mu}$ and  $S_{\bar{\nu}\mu}$ so as we shall see in the next section, using the information on spectra will considerably boost the sensitivity to the new physics. However, the spectra $S_{\nu\mu}$ and $S_{\bar{\nu}\mu}$ are very close to each other and cannot be distinguished. This is due to the fact that $F_{\nu_\mu}^\pi$
and $F_{\bar{\nu}_\mu}^\pi$ are almost equal to each other; see Eqs (\ref{S2},\ref{S3}).  If an excess of $\tau+\bar{\tau}$ is discovered, it will not be possible to distinguish if it comes from the lepton number conserving $\pi^+\to \mu^+ \nu_\tau$ process or from the  lepton number violating 
$\pi^+\to \mu^+ \bar{\nu}_\tau$ process at FASER$\nu$ by studying the energy  spectrum of the events. One suggestion is to attune Q1-3 quadrapole and D1 dipole (located close to the interaction point) such that the transverse distribution of neutrinos emitted from $\pi^+$ and $\pi^-$ decays can be distinguished from one another.

The uncertainties in the predictions of the fluxes of $\nu_\mu$, $\bar{\nu}_\mu$, $\nu_e$ and  $\bar{\nu}_e$ are relatively small but the predictions for $\nu_\tau$ and $\bar{\nu}_\tau$ suffer from large uncertainties. Ref.~\cite{Kling:2021gos} shows that the different simulators  predict the $\nu_\tau$ flux which can differ from each other by more than 100 \%. 
To draw the background  $\tau+\bar{\tau}$ spectrum, we have used $F(\nu_\tau)$ and $F(\bar{\nu}_\tau)$ from a simulator whose prediction is close to the median of the predictions of other  simulators and is therefore recommended by Ref.~\cite{Kling:2021gos}. More details  are described in the end of sect.~\ref{FP}. 
 We shall show in  sect.~\ref{signature} that  the number of events from new physics at FASER$\nu$ will be too low to reconstruct the spectra but, at FASER$\nu$2 with about 400 times more statistics, reconstructing the spectra of the events from new physics may become possible. By then, more dedicated simulations can reduce uncertainties in the $\nu_\tau$ and $\bar{\nu}_\tau$ flux predictions. Moreover, as we discuss in the next section, the data from FASER$\nu$ during the run III of the LHC  can itself determine  which simulator for the  $\nu_\tau$ and $\bar{\nu}_\tau$ fluxes is valid. Thus, before the start of high luminosity run of the LHC and FASER$\nu$2 data taking, the uncertainty in the standard model prediction for $F(\nu_\tau)$ and $F(\bar{\nu}_\tau)$ can be significantly reduced. 
\section{Characteristics of FASER$\nu$, SND@LHC and FASER$\nu$2 \label{FP}}
The FASER$\nu$  and SND@LHC detectors are  respectively located in the side tunnels TI12 and TI18, 480 m  downstream the ATLAS Interaction Point (IP). FASER$\nu$ is composed of 1000 emulsion layers interleaved with 1 mm tungsten plates \cite{Abreu:2019yak}. The effective masses of FASER$\nu$ \cite{Abreu:2019yak} and SND@LHC  \cite{SND} are respectively 1.2 ton and 800 kg and their sizes are $25 ~{\rm cm}\times 25 ~{\rm cm}\times 1.3~{\rm m}$ and $41.6 ~{\rm cm}\times 38.7 ~{\rm cm}\times 32~{\rm cm}$, respectively. Both detectors boast having excellent spatial and angular resolution in reconstructing the tracks of charged particles which will enable them to resolve the $\nu_\tau$ CC events.  The details of the FASER$\nu$ and SND@LHC detectors are presented in \cite{Abreu:2019yak} and in \cite{SND}, respectively. An updated prediction for the fluxes at these detectors can be found in \cite{Kling:2021gos}.
The upgrade of FASER$\nu$
 for the high luminosity LHC will have a size of  $40 ~{\rm cm}\times 40 ~{\rm cm}\times 8~{\rm m}$ and a mass of  20 tonnes \cite{Anchordoqui:2021ghd}. The proposed location of this detector could be slightly off-axis at a distance of 620 meter or on-axis at a distance of 480 meter from the interaction point \cite{Anchordoqui:2021ghd}. Ref.~\cite{Kling:2021gos} has also  predicted the neutrino
  within the SM at this detector which is assumed 
  to be placed 620 m downstream from the ATLAS IP.
 \footnote{\texttt{https://github.com/KlingFelix/FastNeutrinoFluxSimulation}}
 
According to \cite{Aad:2019ugc}, the efficiency of FASER$\nu$ in detecting 1-prong $\tau$ decay is 75 \% and that of 3-prong decay is 15 \%. Considering that the branching ratios of 1-prong and 3-prong are respectively 85 \% and 15 \%, we take the average efficiency of 67 \% for the $\nu_\tau$ detection at FASER$\nu$.
We take similar efficiency for the tau neutrino detection at SND@LHC. 
Considering table II of \cite{Abreu:2019yak}, throughout our analysis, unless it is stated otherwise, we take 15 \% uncertainty in the neutrino flux normalization. 

It is shown in Ref.~\cite{Abreu:2019yak} that the resolution of the $\nu_\mu$ energy measurement will be 30\%. However, to our best knowledge, similar analysis has not been carried out  for the energy resolution of $\nu_\tau$ and $\nu_e$ or for the detected $\tau$.  The tau particles at forward experiments will travel a distance of $L_{tr}\sim 1~{\rm cm} (E_\tau/200~{\rm GeV})$ before decay. The direction of the momentum of $\tau$ (or equivalently, the direction of the line connecting decay and production vertices) can be reconstructed with a precision of $0.06 (1~{\rm cm}/L_{tr})$ mrad \cite{Abbasi:2020zmr}. Due to the lepton flavor conservation, all decay modes of $\tau$ contain $\nu_\tau$ which appears as missing energy momentum.
In hadronic decay modes of $\tau$, which constitute 65\% of the decays, only one neutrino is emitted. By measuring the energy-momentums of visible particles and reconstructing the direction of $\tau$ momentum and using energy-momentum conservation, it will be therefore possible to reconstruct the $\tau$ energy for the hadronic decay modes. For example, in case of $\tau^+\to \pi^+ \nu_\tau$, the energy of the final pion and the angle that it makes with the direction of the tau momentum determine the energy of the $\tau$. Thanks to the sub-milliradian  angular resolution of FASER$\nu$, the tiny angle  [$O(10^{-2})$] between  the directions of tau and pion momenta can be measured with remarkable accuracy so the energy resolution in determining the tau energy will mainly be limited by the energy resolution in measuring the energy of $\pi^+$. We take nominal value of  30 \% for the $\tau$ energy reconstruction when necessary. As we shall see in the next session, at FASER$\nu$ and SND@LHC during the run III of the LHC, the maximum signal events will be too small to justify binning the data so we shall only analyze the total number of predicted $\tau$ events  for the run III in studying the sensitivity for new physics.  Of course without binning the data,  the tau energy measurement will not be relevant.

\begin {table}
\caption { {The number of $\tau$ events within the SM at FASER$\nu$ using the prediction of three different simulators \cite{Kling:2021gos}. The relative $\chi^2$ minimized over normalization uncertainty of 15 \%  is shown in the last column, assuming  Pythia8 (Hard)  as the true model.}}\label{tab:Flux_tau} 
\begin{center} 
\end{center}
  \begin{center}
  	\begin{tabular} {| c | c| c| c|c|c|c|}
  	\hline
  	\multirow{2}{*}{Simulator} & \multicolumn{5}{c|}{bin limits in GeV} &  \multirow{2}{*}{$\chi^2_{rel}$}   \\
  	
  		\cline{2-6} & $< 50  $ & $50 - 100$ & $100 - 500$ &$500 - 1000$ & $1000< $& \cr  \hline
  		
  		Pythia8 (Hard)&  0.9 & 1.8 & 8.1  &9.7  &4.8   & 0.0 \cr
  		DPMJET 3.2017&  1.5  & 3.1 & 16.2 &23.3 &14.5&43.7 \cr
  		SIBYLL 2.3c&  0.7   & 1.1 & 3.7   &3.1  &0.7   &9.6 \cr
  		\hline
  	\end{tabular}       
  \end{center}
\end{table}

 As is well-known, the predictions of different simulators for  the $\nu_\tau$ and $\bar{\nu}_\tau$ spectra at the forward experiments are significantly different.  The prediction of  simulator, $J$ for the  number of  $\tau$ events at the $i$th bin  can be written as 
 	\begin{eqnarray} {B}_i^J= &\epsilon_\tau N_W \int_{E^i_{min}}^{E^i_{max}}
 	\int_{m_\tau}\int_{E_\tau} \Bigl[ F_{\nu_\tau}^J(E_\nu) 
 	\frac{d\sigma_{CC}}{dE_\tau}(\nu_\tau +{\rm nucleus}\to \tau +X)+  \nonumber \\  &F_{\bar{\nu}_\tau}^J(E_\nu)  \frac{d\sigma_{CC}}{dE_\tau}(\bar{\nu}_\tau +{\rm nucleus}\to \tau^+ +X) \Bigr] f(E'_\tau,E_\tau)  dE_\nu dE_\tau dE'_\tau \label{Bis} ,\end{eqnarray}
 where $F^J_{\nu_\tau}$ and $F^J_{\nu_\tau}$ are respectively $\nu_\tau$ and $\bar{\nu}_\tau$ energy spectra predicted  by different simulators in Ref. \cite{Kling:2021gos}.  The superscript $J$ determines the simulator. $\epsilon_\tau=0.67$ is the efficiency of the $\nu_\tau$ detection at the  detector. $f(E_\tau',E_\tau)$ is the energy resolution function which we take to be a Gaussian with a 30 \% width. $(E_{min}^i,E_{max}^i)$ determine the limits of the $i$th energy bin.
 $N_W$ is the number of tungsten nuclei inside the detector, $N_W=M_D/M_W$ where  $M_W=183 m_p$ and $M_D=1.2$ ton for run III detector.  In table \ref{tab:Flux_tau}, we show  our prediction for the number of events in different energy bins at FASER$\nu$.
  
   Taking the uncertainty on the flux normalization to be $\sigma_\eta=15 \%$, we have computed $\chi^2_{rel}$ as defined below for each model, $J$, and minimized over the pull parameter, $f$:
 \be \chi^2_{rel}=\sum_i \left[ \frac{[(1+f)B_i^{true}-N_i^J]^2}{B_i^{true}}+\frac{f^2}{\sigma_\eta^2} \right]. \ee
 In computing $\chi_{rel}^2$ that is shown in last column of table  \ref{tab:Flux_tau}, we take $B_i^{true}$ to be equal to the prediction of Pythia8 (Hard). Notice that since our bin sizes are large, the results should be robust against the value of the energy resolution.
 As seen in table \ref{tab:Flux_tau}, the FASER$\nu$ experiment can discriminate between different simulators predicting the $\nu_\tau$ flux with high confidence level. We therefore assume a well-known shape of the fluxes predicted within the SM in making forecast  for FASER$\nu$2. We shall however study the impact of  the normalization uncertainty. 
 
Hereafter in this study, for computing the SM background for the $\tau+\bar{\tau}$ events, we take the predictions of Pythia8 (Hard) given in  Ref. \cite{Kling:2021gos} for $F_{\nu_\tau}$ and $F_{\bar{\nu}_\tau}$. The   Pythia8 (Hard) prediction is close to the median of the predictions of  the DPMJET 3.2017 and SIBYLL 2.3c simulators.  The other input spectra that we require for our computations are the fluxes of $\nu_\mu$ and   $\bar{\nu}_\mu$ that are  sourced by charged  pion mesons, $F_{\nu_\mu}^\pi$ and $F_{\bar{\nu}_\mu}^\pi$.  Fortunately, the differences between the predictions by different simulators  for $F_{\nu_\mu}^\pi$ and $F_{\bar{\nu}_\mu}^\pi$ are negligible. For our computations, we use the average of the predictions given in Ref.~\cite{Kling:2021gos}.
 
\section{Signatures of the models in  Forward Experiments \label{signature}}
In this section, we study how FASER$\nu$  and  SND@LHC during LHC run III  can constrain  the models introduced in sect. \ref{model}. We compare the bounds with the results of \cite{Falkowski:2021bkq} which has performed a similar analysis within the framework of effective field theory. We also compare our bounds with the existing bounds and the one to be derived by upcoming DUNE experiment  
\cite{Giarnetti:2020bmf}. 
We then show how FASER$\nu$2 can improve the results with or without reconstructing the energy spectrum of $\tau$.


\begin{center} {A. $\pi^+ \to \mu^+ \nu_\tau$}
\end{center}

Similarly to Eq. (\ref{Bis}), we compute the number of signal events per bin as follows
\begin{eqnarray} \mathcal{N}_s^i= &\epsilon_\tau N_W Br(\pi^+ \to \mu^+ \nu_\tau)\int_{E^i_{min}}^{E^i_{max}}
\int_{m_\tau}\int_{E_\tau} \Bigl[ F_{\nu_\mu}^\pi(E_\nu)  
\frac{d\sigma_{CC}}{dE_\tau}(\nu_\tau +{\rm nucleus}\to \tau +X)+  \nonumber \\  &F_{\bar{\nu}_\mu}^\pi(E_\nu)  \frac{d\sigma_{CC}}{dE_\tau}(\bar{\nu}_\tau +{\rm nucleus}\to \tau^+ +X) \Bigr] f(E'_\tau,E_\tau)  dE_\nu dE_\tau dE'_\tau \label{Nsi} \end{eqnarray}
To compute the SM background per bin, $B_i$, we use Eq.~(\ref{Bis}). 
The total observed number in bin $i$ is then $N_i^{obs}=B_i +\mathcal{N}_s^i$. We define the $\chi^2$ as follows
\be \chi^2 =\sum_i \left[ \frac{\left( B_i (1+\eta) -N_i^{obs}\right)^2}{B_i} + \frac{\eta^2}{\sigma_\eta^2} \right] \ee
where $\eta$ is the pull parameter that takes care of the uncertainty in normalization,  mainly coming from the uncertainty in the cross section and the flux normalization,  $\sigma_\eta=15\%$.

\begin {table}
\caption { Total expected number of $\tau$ events at FASER$\nu$ and SND during the run III and at FASER$\nu$2 during the high luminosity run of the LHC. To compare the values, we have saturated the bounds on the new physics, setting $G_e$ equal to $5 \times 10^{-7}$ GeV$^{-2}$ and the branching ratios to $2.4 \times 10^{-3}$. The last column shows the SM background computed using fluxes described in the last paragraph of  Sec.~\ref{FP} }\label{tab:TOTALtau} 
\begin{center} 
\end{center}
  \begin{center}
  	\begin{tabular} {| l | c| c| c| c|}
  		\hline
  		Detector &  $Br(\pi^+\to \nu_\tau\mu^+)$ & $Br(\pi^+\to \bar{\nu}_\tau\mu^+)$&$G_e$ 
  		&  SM \cr  \hline
  		\hline
  		SND@LHC&  1.0 &0.9&0.003& 6.6
  		\cr
  		FASER$\nu$ &  4.9& 4.3 &0.027 & 25.3 
  		\cr
  		FASER$\nu$ 3.6 &  1125.9 & 938.0 & 9.6 & 3403.3 \cr
  		\hline
  	\end{tabular}       
  \end{center}
\end{table}

  The second column in table \ref{tab:TOTALtau} shows total $\tau$ events ($\sum_i \mathcal{N}_s^i$) originated from $\pi^+\to \nu_\tau \mu^+$ with a branching ratio saturating the present bound. As seen, the number of events at FASER$\nu$ and SND@LHC during the run III of the LHC data cannot reach  a statistical limit so binning the data  does not make sense. To compute $\chi^2$ for these experiments, we consider only one bin ({\it i.e.,} the total events).   However, at FASER$\nu$2 during the high luminosity  run the statistics will be large enough to use binning.

 \begin{figure}
 	\centering
 	\includegraphics[width=0.7\textwidth]{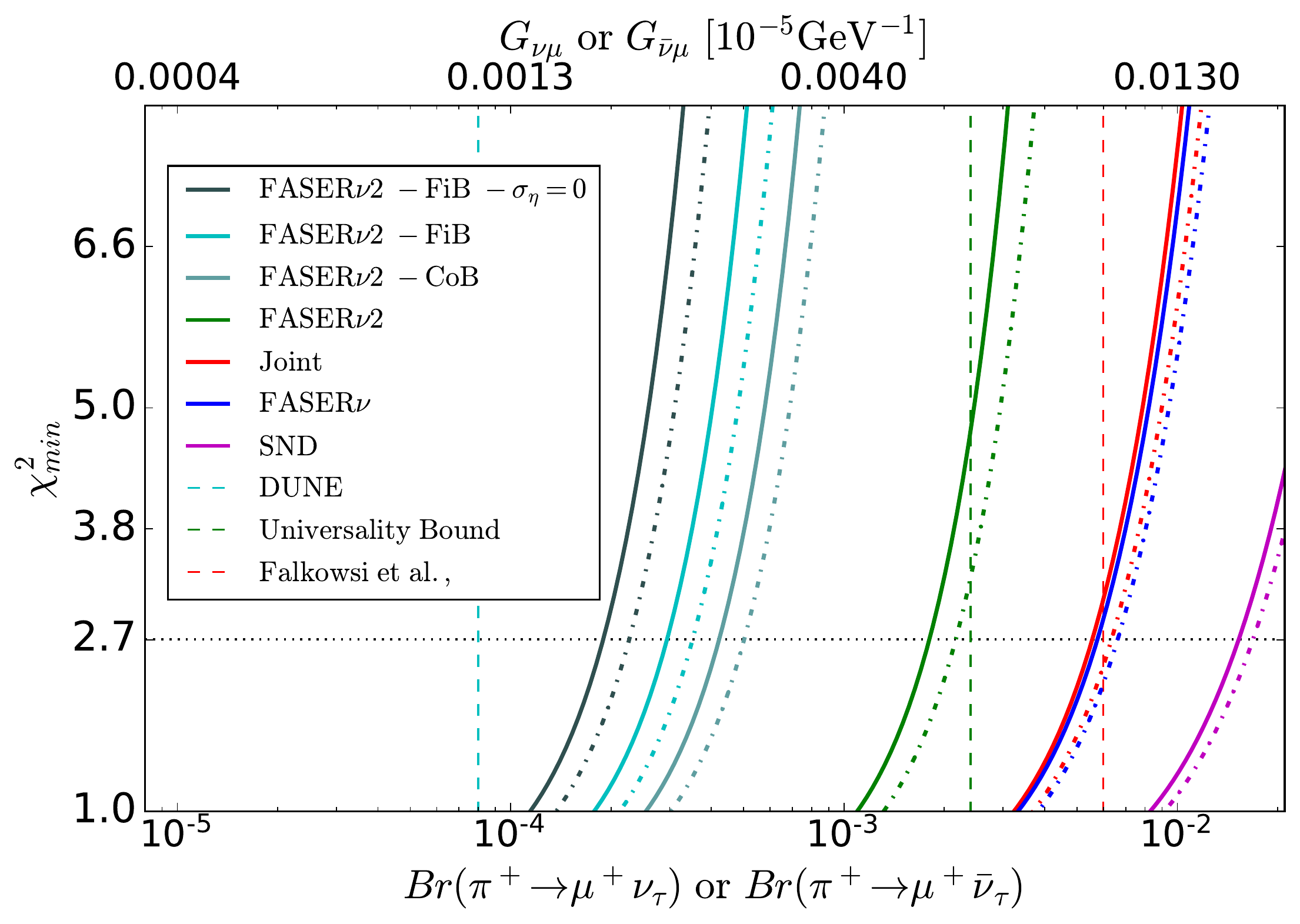}
 	\caption{ $\chi^2$ versus $Br(\pi^+\to \mu^+\nu_\tau)$ (solid curves) or  versus $Br(\pi^+\to \mu^+\bar{\nu}_\tau)$ (dash-dotted curves).  Purple, blue and green curves respectively correspond to SND@LHC, FASER$\nu$ and FASER$\nu$2 without binning. The red curve shows the $\chi^2$ for  SND@LHC and FASER$\nu$ combined. The curves marked with ``FiB" and ``CoB" show $\chi^2$ for FASER$\nu$2 with an energy resolution of 30 \%  and with two different binning patterns corresponding to fine and coarse binning. More detail are described in the text. Drawing all these curves, except the black one(s), we have assumed a 15 \% uncertainty in the normalization of the flux and have minimized over the corresponding pull parameter. Drawing the black curve(s), we have assumed zero uncertainty in the flux normalization and fine binning. The vertical lines from left to right correspond to the expected DUNE bound  \cite{Giarnetti:2020bmf}, the bound 
 from the lepton flavor universality  constraint \cite{Aguilar-Arevalo:2015cdf} and the  forecast by \cite{Falkowski:2021bkq} for FASER$\nu$.} 	\label{fig:chis}
 \end{figure}

 The solid lines in  Fig.~\ref{fig:chis} show the $\chi^2$ minimalized over $\eta$ (the normalization uncertainty). For the solid curves, the horizontal axis is $Br(\pi^+ \to \mu^+ \nu_\tau)$ which in terms of $G_{\nu \mu}$ can be written as 
  $Br(\pi^+ \to \mu^+ \nu_\tau)=(G_{\nu\mu}^2/8G_F^2)[m_\pi^4/(m_\mu^2 (m_u+m_d)^2)]$.  
 The upper horizontal axis shows the corresponding effective coupling.
 The purple, blue and red curves show $\chi^2$ minimized over the flux normalization uncertainty for SND@LHC, FASER$\nu$ and their combination, respectively. Since the normalization uncertainty mainly originates from the production rate at the interaction point which is common for SND@LHC and FASER$\nu$, we treat the uncertainty with a single pull parameter when combining the SND@LHC and FASER$\nu$ predictions. For FASER$\nu$2, we have used three different binning schemes:  (1) no binning; (2) coarse binning with bins divided as $E_\tau<50$ GeV, $50~{\rm GeV}< E_\tau<100$~GeV, $100~{\rm GeV}< E_\tau<500$~GeV, 
 $500~{\rm GeV}< E_\tau<1$~TeV and $1~{\rm TeV}<E_\tau$;
(3) fine binning  with  three bins at each energy decade.
 Drawing all these curves, we have taken an energy resolution of 30 \% in the $E_\tau$ determination. Of course, for no binning case, the results do not depend on the energy resolution. Even for the coarse binning,  $\chi^2$  is robust against varying  the energy resolution. In all curves, except for the black one(s), we have taken the flux normalization uncertainty equal to 15 \%. Comparing the FASER$\nu$2 curves with each other, it is clear that binning the data (or in other words, using the spectral information) dramatically increases the sensitivity to the new physics signal from $\pi^+ \to \mu^+ \stackrel{(-)}{\nu}_\tau$. Studying Fig.~\ref{fig:spectrum}, this is understandable as the spectral shape of the background is considerably  harder than the signal spectrum so the new physics signal cannot be hidden in the normalization uncertainty once the spectral uncertainty has been taken into account. Comparing black curve(s) with the rest, we observe that the uncertainty in the normalization significantly reduces the sensitivity.

Notice that for computing $\chi^2$, we have set the ``observed" number of events per bin equal to the ``average" background plus the ``average" predicted signal for ``true" Br$(\pi^+\to \mu^+\stackrel{(-)}{\nu}_\tau$) rather than having real data.  Thus, if we want to minimize $\chi^2$ over the only free parameter of the model which is Br$(\pi^+\to \mu^+\stackrel{(-)}{\nu}_\tau$) we will invariably obtain
 zero. In fact, the $\chi^2$ that we are computing will have a $\chi^2$ distribution with only 1 (=number of pull parameters) degrees of freedom. As a results, regardless of the number of bins,  the horizontal line at 2.7 represents 90 \% C.L. For example, we find that  FASER$\nu$2 with coarse binning can constrain $Br(\pi^+ \to\mu^+\nu_\tau)<4\times 10^{-4}$ at 90 \% C.L.
 
 The vertical line at $6\times 10^{-3}$  is the forecast at 90 \% for FASER$\nu$ found by \cite{Falkowski:2021bkq} which is in qualitative agreement with our results.  \footnote{ In the notation of \cite{Falkowski:2021bkq}, $G_{\nu \mu}=2 (V_{CKM})_{11}(\epsilon_P)_{\mu \tau}/v^2$.} The vertical line at $2.4 \times 10^{-3}$ shows the present bound from the flavor universality of $\pi^+$ decay \cite{Aguilar-Arevalo:2015cdf}.   Finally the forecast for DUNE near detector sensitivity at 90 \% is shown as a vertical line at  $8\times 10^{-5}$ \cite{Giarnetti:2020bmf}.
 As seen from the figure, FASER$\nu$ and SND@LHC will not be able to improve the present bounds but FASER$\nu$2 will have a good prospect of improving the bound or find a signal. Reconstructing the energy spectrum of $\tau$ with a moderate energy resolution 
can dramatically improve the reach for new physics.

 The effects appearing for $\pi^+ \to \nu_\tau \mu^+$ can be reinterpreted for a variety of other models, too. For example, let us consider a model that leads to the decay of $\pi^+$ to $\mu^+$ and a sterile neutrino $\nu_s$: $\pi^+ \to \mu^+ \nu_s$. As long as $\nu_s$ is lighter than $\sim 1$ MeV, the bound from flavor universality of the pion decay applies for this decay mode, too: $Br(\pi^+ \to \mu^+ \nu_s)<2.4 \times 10^{-3}$. In principle, $\nu_s$ can have a mixing as large as $|U_{\tau 4}|\sim 0.3$ with $\nu_\tau$. Such scenario is motivated by the two anomalous $\nu_\tau$ events observed by ANITA  \cite{Cherry:2018rxj,Farzan:2021gbx}. If the mass of $\nu_s$ is  much smaller than $20~{\rm eV} (E_\nu/{\rm TeV})^{1/2}(480~{\rm m}/L)^{1/2}$, oscillation will not take place before reaching the FASER$\nu$ detector so 
 we would not have any excess due to $\nu_s \to \nu_\tau$. With a sterile neutrino mass of $\sim 20$ eV, there will be a $\tau$ excess with oscillatory behavior with $E_\nu$. For sterile neutrino mass much larger than 20~eV, the $\nu_s\to \nu_e$ oscillation probability will average to $2|U_{\tau 4}|^2(1-|U_{\tau 4}|^2)$ so there will be a $\tau$ excess with similar spectrum as that from $\pi^+ \to \nu_\tau \mu^+$. The discussion and results on $\pi^+ \to \nu_\tau  \mu^+$ also applies for this case,
 replacing $Br(\pi^+ \to \nu_\tau \mu^+)$ with $2|U_{\tau 4}|^2(1-|U_{\tau 4}|^2)Br(\pi^+ \to \nu_s \mu^+)$. Notice that in the scenario described in this paragraph, unlike the canonical $3+1$ scheme, the sterile neutrino is produced by new physics (e.g., an intermediate new scalar) at pion decay rather than by oscillation. For a study of 3+1 scheme for the FASER$\nu$ detector, see \cite{Bai:2020ukz}.
 
 \begin{center} {B. $\pi^+ \to \mu^+ \bar{\nu}_\tau$}
 \end{center}
 The number of signal events in this case is given by Eq. (\ref{Nsi}) by replacing $Br(\pi^+ \to \mu^+ {\nu}_\tau)$ with $Br(\pi^+ \to \mu^+ \bar{\nu}_\tau)$ and swapping $F_{\bar{\nu}_\mu}^\pi(E_\nu)$ with $F_{{\nu}_\mu}^\pi(E_\nu)$ as follows
\begin{eqnarray} \mathcal{{N}}_s^i= &\epsilon_\tau N_W Br(\pi^+ \to \mu^+ \bar{\nu}_\tau)\int_{E^i_{min}}^{E^i_{max}}
\int_{m_\tau}\int_{E_\tau} [ F_{\bar{\nu}_\mu}^\pi(E_\nu)  
\frac{d\sigma_{CC}}{dE_\tau}(\nu_\tau +{\rm nucleus}\to \tau +X)+ \nonumber \\  &F_{{\nu}_\mu}^\pi(E_\nu)  \frac{d\sigma_{CC}}{dE_\tau}(\bar{\nu}_\tau +{\rm nucleus}\to \tau^+ +X)]f(E'_\tau,E_\tau)  dE_\nu dE_\tau dE'_\tau \label{NsiBAR} ,\end{eqnarray}
The total number of signal events (summed over all bins) are shown in the third column of table \ref{tab:TOTALtau} for $Br(\pi^+ \to \mu^+ \bar{\nu}_\tau)$ saturating the bound from universality measurement \cite{Aguilar-Arevalo:2015cdf}.
 The corresponding $\chi^2$ is also shown in Fig.~\ref{fig:chis} with {dash-dotted} line. The bound at $2.4 \times 10^{-3}$ from the flavor universality of the pion decay applies for this model, too. Since $F_{\nu_\mu}\simeq F_{\bar{\nu}_\mu}$, the curves corresponding to $Br(\pi^+ \to \mu^+ \nu_\tau)$ and $Br(\pi^+ \to \mu^+ \bar{\nu}_\tau)$ are very close to each other.
 
 \begin{center} {C. $\nu_e +{\rm nucleus}\to \tau +X$}
 \end{center}

 Let us now assess the effects of $G_e$ and the $\nu_e +{\rm nucleus}\to \tau +X$ process introduced in sect. \ref{TauProd}. The excess in $\tau$ events in this case can be written as
\begin{eqnarray} \mathcal{{N}}_s^i= &\epsilon_\tau N_W\int_{E^i_{min}}^{E^i_{max}}
\int_{m_\tau}\int_{E_\tau} [ F_{{\nu}_e}(E_\nu)  
\frac{d\sigma_{CC}}{dE_\tau}(\nu_e +{\rm nucleus}\to \tau +X)+ \nonumber \\  &F_{\bar{\nu}_e}(E_\nu)  \frac{d\sigma_{CC}}{dE_\tau}(\bar{\nu}_e +{\rm nucleus}\to \tau^+ +X)]f(E'_\tau,E_\tau)  dE_\nu dE_\tau dE'_\tau \label{NsiBAR} ,\end{eqnarray}
 Equating  $G_e$ with the bound from $\tau^- \to \pi^- \nu_e$ ({\it i.e.,} setting $G_e=5\times 10^{-7}$ GeV$^{-2}$), table \ref{tab:TOTALtau} shows the predicted number  of events at FASER$\nu$, SND@LHC and FASER$\nu$2. Even at FASER$\nu$, the number of the signal events cannot exceed 10.  As a result, we confirm the conclusion of   \cite{Falkowski:2021bkq} that the planned forward experiments cannot improve the bounds on $G_e$. 
 
 
\section{Summary and discussion \label{summary}}
We have shown that the FASER$\nu$ detector will provide a breakthrough on our understanding of the interactions of the third generation leptons. We have focused on three beyond SM LFV processes that can give rise to a $\tau$ excess at the FASER$\nu$ detector:  (i) $\pi^+ \to \mu^+ \nu_\tau$, (ii) $\pi^+ \to \mu^+ \bar{\nu}_\tau$
and (iii) $\nu_e +{\rm nucleus}\to \tau +X$. We have introduced three models based on adding new scalars charged under electroweak symmetry that give rise to these processes. We have shown that by imposing   proper  global $U(1)$  flavor symmetries, the desired flavor patterns of the Yukawa couplings between the SM fermions can be explained. The same symmetry can also explain the smallness of the masses of the first generation fermions of the SM.

 Our model for $\pi^+\to \mu^+ \nu_\tau$ 
contains two new scalar doublets coupled respectively to the leptons and the quarks. In the  presence of a mixing between the charged components of the two doublets, integrating out the heavy intermediate states we  obtain a pseudoscalar-scalar four Fermi effective coupling, $G_{\nu \mu}$, shown in Eq. (\ref{eff-pi-mu}) which gives rise to $\pi^+ \to \mu^+ \nu_\tau$ with a rate enhanced by $m_\pi^2/(m_u+m_d)^2$. The bound on the deviation of $\Gamma (\pi^+ \to e^+ \nu)/\Gamma (\pi^+ \to \mu^+ \nu)$  from the standard model prediction \cite{Aguilar-Arevalo:2015cdf} then implies $G_{\nu \mu}<4 \times 10^{-8}~{\rm GeV}^{-2}$. The $G_{\nu \mu}$ coupling can also lead to $\nu_\tau +{\rm nucleus} \to \mu^+ +X$ in the neutrino scattering experiments but without the $m_\pi^2/(m_u+m_d)^2$ enhancement. The bound on $G_{\nu \mu}$ renders this effect negligible while having $Br(\pi^+\to \nu_\tau \mu^+)\sim 10^{-3}$ which can lead to a sizable $\tau$ excess in the forward experiments.  We have shown that in terms of the formalism developed to study the effects of charged current non-standard interaction on neutrino experiments, we can write $\epsilon^s_{\mu\tau}=[Br(\pi^+ \to \nu_\tau \mu^+)]^{1/2}$ so thanks to  the $m_\pi^2/(m_u+m_d)^2$ enhancement, $\epsilon^s_{\mu\tau}$ can be as large as $O({\rm few}
\times 10^{-2})$. We have argued that bounds from NOMAD on the $\tau$  production do not constrain this model because the energy of the neutrino flux produced by the pion (rather than the Kaon) decay in the NOMAD experiment is too low to lead to the $\tau$ production. In our model, the bounds from $\tau^+ \to \mu^+ \pi^0$ can be satisfied because the neutral components of the new scalar doublets do not mix. We have pointed out that the model can also explain the observed $(g-2)_\mu$ deviation from the SM prediction.

By changing the flavor $U(1)$ charge assignment to the leptons in the model described above, we obtain a model giving rise to the $G_e$ effective coupling in Eq. (\ref{GeGmu}) which leads to $\nu_e +{\rm nucleus}\to \tau +X$. The $G_e$ coupling yields LFV decay mode $\tau^+ \to \pi^+\nu$ with a rate enhanced by $m_\pi^2/(m_u+m_d)^2$.  The strong bound on this decay mode severely constrains $G_e$. The model for the lepton number violating $\pi^+ \to \mu^+ \bar{\nu}_\tau$ is based on introducing a scalar doublet coupled to the  quarks plus a singlet charged scalar with an off-diagonal coupling to left-handed doublets of the second and third generations. Such a singlet is also motivated with a small anomaly observed in the $\tau$ decay \cite{Crivellin:2020klg,Amhis:2019ckw}. We find that in the same range of the parameter space that can explain this deviation from the SM prediction, a sizable rate of $\pi^- \to \mu^- \nu_\tau$ (leading to discernible tau excess in the forward experiments) can be obtained.

The bounds on the effective coupling discussed from the  (lepton+ missing energy) 
signal at the LHC \cite{Falkowski:2021bkq,Miss} do not apply  for our models because the 
new intermediate states whose integrating out lead to these effective couplings have mass around $O$(300 GeV) which is close to the center of mass energy of the partons scattering at the LHC. That is at the LHC, we cannot  
use the effective coupling formalism to describe the effects of new physics described in the present paper. We have briefly  discussed the production of the new scalars and their potential signals at CMS and ATLAS \cite{Majid}. 

We have then studied the potential of forward experiments to test these models by looking for the $\tau$ event excess. The bound forecasts for FASER$\nu$ and SND@LHC by the present work  is in agreement with those found  in Ref.~\cite{Falkowski:2021bkq}. We have proceeded by studying the energy spectrum of the tau events and showed that since the energy spectrum of the signal is going to be considerably softer than the $\tau$ event background, constructing the energy spectrum at FASER$\nu$2 can significantly improve the sensitivity to the new physics. We have discussed the possible resolution of the $\tau$ energy reconstruction  at FASER$\nu$2 and demonstrated the dependence of sensitivity to the new physics signal on the energy binning scheme.  
 \section*{Acknowledgment} YF  thanks  K. Azizi, M. Hashemi, J. Kopp and  E. Fernandez-Martinez  for useful information and discussion. She is especially grateful to S. Su  and P. Bakhti for the encouragement and for the collaboration in the early stages of this work.
 This project has received funding /support from the European Union’s Horizon 2020 research and innovation programme under the Marie Skłodowska -Curie grant agreement No 860881-HIDDeN. YF has received  financial support from Saramadan under contract No.~ISEF/M/400279 and No.~ISEF/M/99169.
SA is supported by a grant from Basic Sciences Research Fund (No. BSRF-phys-399-01).

\end{document}